\documentclass[twoside]{article}
\usepackage{fleqn,espcrc2,epsf}

\usepackage{graphicx}
\usepackage[figuresright]{rotating}

\newcommand{\AmS}{{\protect\the\textfont2
  A\kern-.1667em\lower.5ex\hbox{M}\kern-.125emS}}

\title{Unquenching the Topological Susceptibility with an Overlap Action}

\author{Tam\'as G. Kov\'acs\thanks{Supported by the European 
Community's Human Potential Programme under contract 
HPRN-CT-2000-00145, ``Hadrons/Lattice QCD'' and 
by Hungarian science grant OTKA-T032501.}\thanks{On leave
from Department of Theoretical Physics, University of P\'ecs,
Hungary.}
            \address{NIC/DESY Zeuthen, Platanenallee 6, 
                     D-15738 Zeuthen, Germany}} 
       
\begin{document}

\begin{abstract}
We estimate the quark-mass dependence of the topological susceptibility
with dynamical overlap and clover fermions. Unquenching effects on the
susceptibility turn out to be well approximated by a reweighting of
a quenched ensemble with a low-eigenmode truncation of the fermionic
determinant. We find that it is most likely due to the explicit 
chiral symmetry breaking of the fermion action that present day
dynamical simulations do not show the expected suppression of the
topological susceptibility.
\vspace{1pc}
\end{abstract}

\maketitle

One of the most profound effects the inclusion of dynamical fermions 
is expected to have on the QCD vacuum is the suppression of
fluctuations of the topological charge. Due to the index theorem in
the continuum, the Dirac operator has $|Q|$ zero modes in the charge
$Q$ sector. In the presence of $N_f$ degenerate flavours of light dynamical
quarks, in the charge $Q$ sector the fermionic determinant is
proportional to $m^{|Q| N_f}$ (to lowest order in $m$).
Light dynamical fermions are therefore expected to suppress 
higher topological sectors and thus also the topological susceptibility, 
$\chi = \langle Q^2 \rangle /V$.

On a more quantitative level, for small enough quark masses chiral
perturbation theory predicts that
\begin{equation}
   \chi_{LS} = \frac{\Sigma m}{N_f^2} = \frac{f_\pi^2 m_\pi^2}{2 N_f},
      \label{eq:LS}
\end{equation}
where $\Sigma$ is the chiral condensate and $m_\pi$ and $f_\pi$ are the
pion mass and decay constant \cite{LS}. This holds provided the volume
is large enough that chiral symmetry is effectively broken, i.e.\
$x=\Sigma m V \gg 1$.

At the other extreme, i.e.\ for heavy quarks, one expects the susceptibility
to approach a constant, namely the quenched value 
$\chi_{\mbox{\tiny quenched}} = (203 \pm 5$MeV$)^4$ \cite{Anna}.
The interpolation of the susceptibility from light to heavy
quarks has recently been discussed by S.\ D\"urr \cite{Durr}.
He noticed that besides the light dynamical quarks the finite
volume of the box also suppresses fluctuations of the topological
charge, since only a finite number of instantons can be
accomodated in a finite box. Assuming that the two suppression 
mechanisms work independently, D\"urr derived and
essentially unique formula for the susceptibility. In terms of 
the pion mass it reads
\begin{equation}
 \chi(m_\pi) = \left( \frac{2 N_f}{m_\pi^2 f_\pi^2} +
                      \frac{1}{\chi_{\mbox{\tiny quenched}}} \right)^{-1},
\end{equation}
having two free parameters, $f_\pi$ and $\chi_{\mbox{\tiny quenched}}$.
D\"urr also pointed out that --- most
probably due to scaling/chirality violations --- all the
presently available dynamical susceptibility data lie significantly
above this curve. 

We expect that the suppression of the susceptibility depends
crucially on the index theorem which in general does not hold
on the lattice \cite{Hip}. The 
reason for this is that Dirac operators that
explicitly break chiral symmetry (e.g.\ the Wilson operator)
do not have exact zero modes and besides, the very definition
of the topological charge is ambiguous on the lattice. 
It is thus not clear how exactly light dynamical fermions
suppress higher $|Q|$ sectors on the lattice.
The aim of the present work is to check whether improving the 
chirality of the fermion action can have a significant effect on
the topological susceptibility. In what
follows, the topological charge is always defined with my overlap
Dirac operator by counting zero modes. Note that this implies
an exact ``index theorem'' which in fact is a tautology with this 
charge definition. 

Unfortunately chiral fermion actions are still tremendously 
expensive and a full-fledged
dynamical simulation is out of question. It turns out however
that in the present case that might not be needed. A reasonable
estimate of the quark-mass dependence of the susceptibility 
can be obtained by reweighting a quenched ensemble with a low-mode
truncation of the fermionic determinant. Computing the lowest 
$N$ eigenvalues of the Dirac operator $D$, the determinant can
be written as 
\begin{equation}
  \det(D+m) \approx \prod_{k=1}^N (\lambda_k + m) \times \dots,
     \label{eq:det}
\end{equation}
which in turn provides the full quark-mass and $N_f$ dependence
of the topological susceptibility. Notice that while the truncation
(\ref{eq:det}) with $N$ small, does not provide a good approximation to the
full determinant, the topological charge is correlated only with the
low-end of the Dirac spectrum. Therefore, the difference in the
effective action between the various charge sectors can be reliably
estimated even with $N$ reasonably small.

Reweighting is generally a bad procedure since control over 
the errors is lost exponentially fast with increasing volume.
In certain cases, however it can happen that there is still a 
large enough window in the volume where interesting results can be
obtained (see e.g.\ \cite{Fodor} for a recent example). 
In the following I give some reasons why 
reweighting could work for the topological susceptibility.

First of all, there are two sources of
the fluctuations in the fermionic determinant. Eigenvalues $\lambda_k$
with $k>N$, which is by far the largest contribution since with our
choice of $N$, most of the eigenvalues belong to this sector. 
This part does not appear in our approximation. For light quarks
an important contribution to the fluctuations comes from the zero modes
but this is not harmful since this is exactly the quantity we want to
measure: the difference in the effective action between the various charge
sectors. Secondly, as the sea quark-mass changes, in principle, all the
other parameters of the theory (lattice spacing, etc.) change. But
this is not the case if only the low eigenvalues are used for the
reweighting. Clearly e.g. the heavy quark potential is expected to be 
the same in all topological sectors. In spite of all this, the
results that I present here have to be considered as preliminary.
The error bars are estimated with a jack-knife analysis and a more
careful error analysis will appear in a future publication. This is
certainly needed to be able to judge how far one can trust the results
in the small quark mass limit.

The simulations were done with the Wilson plaquette action at $\beta=5.85$
corresponding to a lattice spacing of about $0.12-0.13$fm. The volume
of the box was set to $2.5$fm$^4$ which turned out to be a good compromise;
this is already a physically reasonable volume (if somewhat small) and 
at the same time the reweighting still works reasonably well. The
overlap action that I use is constructed from 10 times APE smeared fat links
\cite{Smear} and the $c_{sw}=1.0$ clover action.
The results are based on 300
configurations and the full run took about 10-20 PC-months.

\begin{figure}[htb!]
\begin{minipage}{75mm}
\resizebox{\textwidth}{!}{
\includegraphics{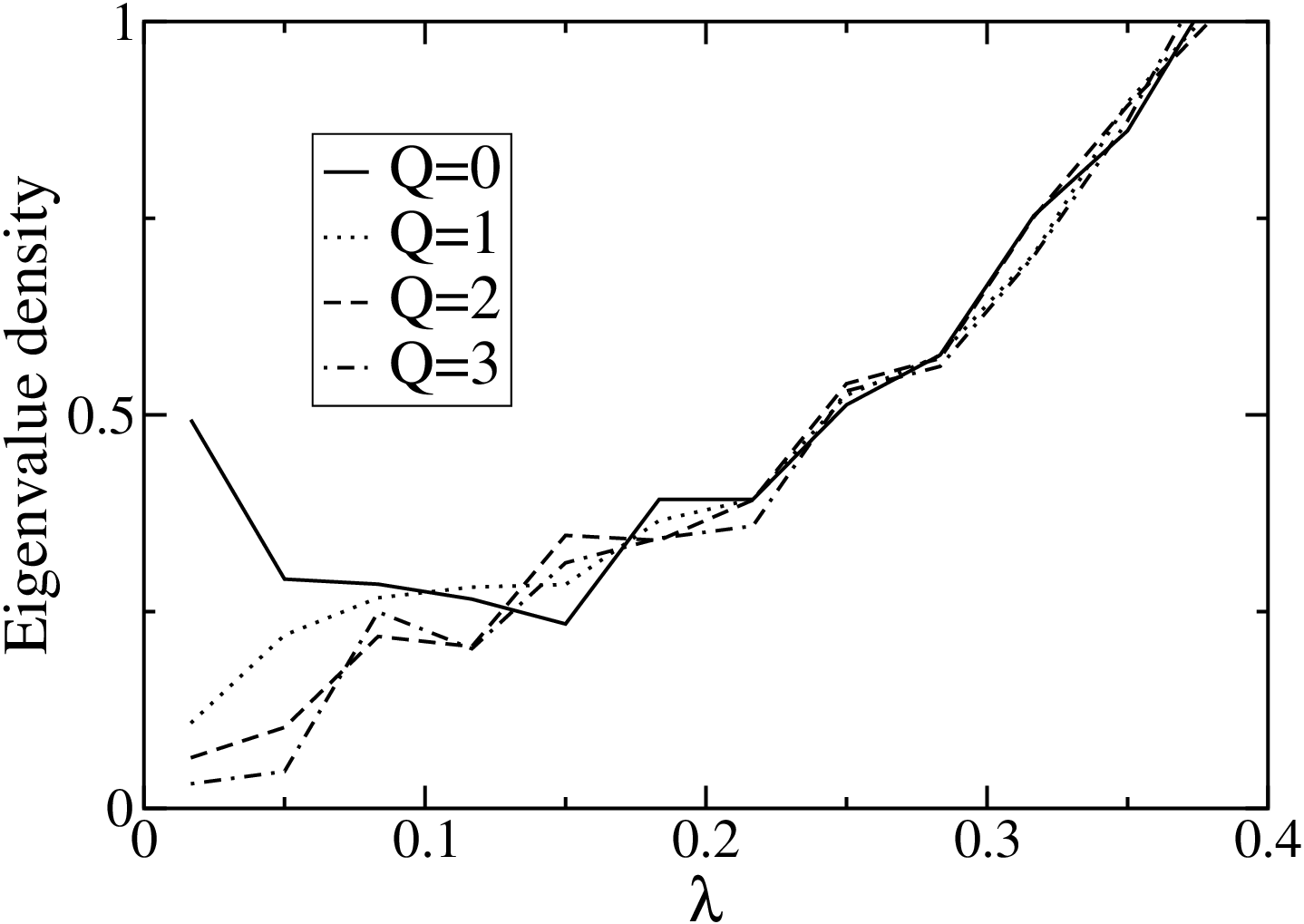}}
\vspace{-1.2cm}\caption{\label{fig:Dspec} The spectral density of the 
overlap Dirac operator in different $Q$ sectors.}
\end{minipage}
\vspace{-3mm}
\end{figure}

The first question that arises is where exactly to truncate the determinant.
To find this out, in Fig.\ \ref{fig:Dspec} I show the spectrum of the
overlap Dirac operator in the different topological charge sectors.  
There does not appear to be a difference between the various $Q$
sectors above roughly $\lambda > 0.20$.  Therefore I truncate the
determinant around $\lambda = 0.2-0.4$ which at the given volume 
translates into taking the lowest $N=10-30$ eigenvalues. 

\begin{figure}[htb!]
\begin{minipage}{75mm}
\resizebox{\textwidth}{!}{
\includegraphics{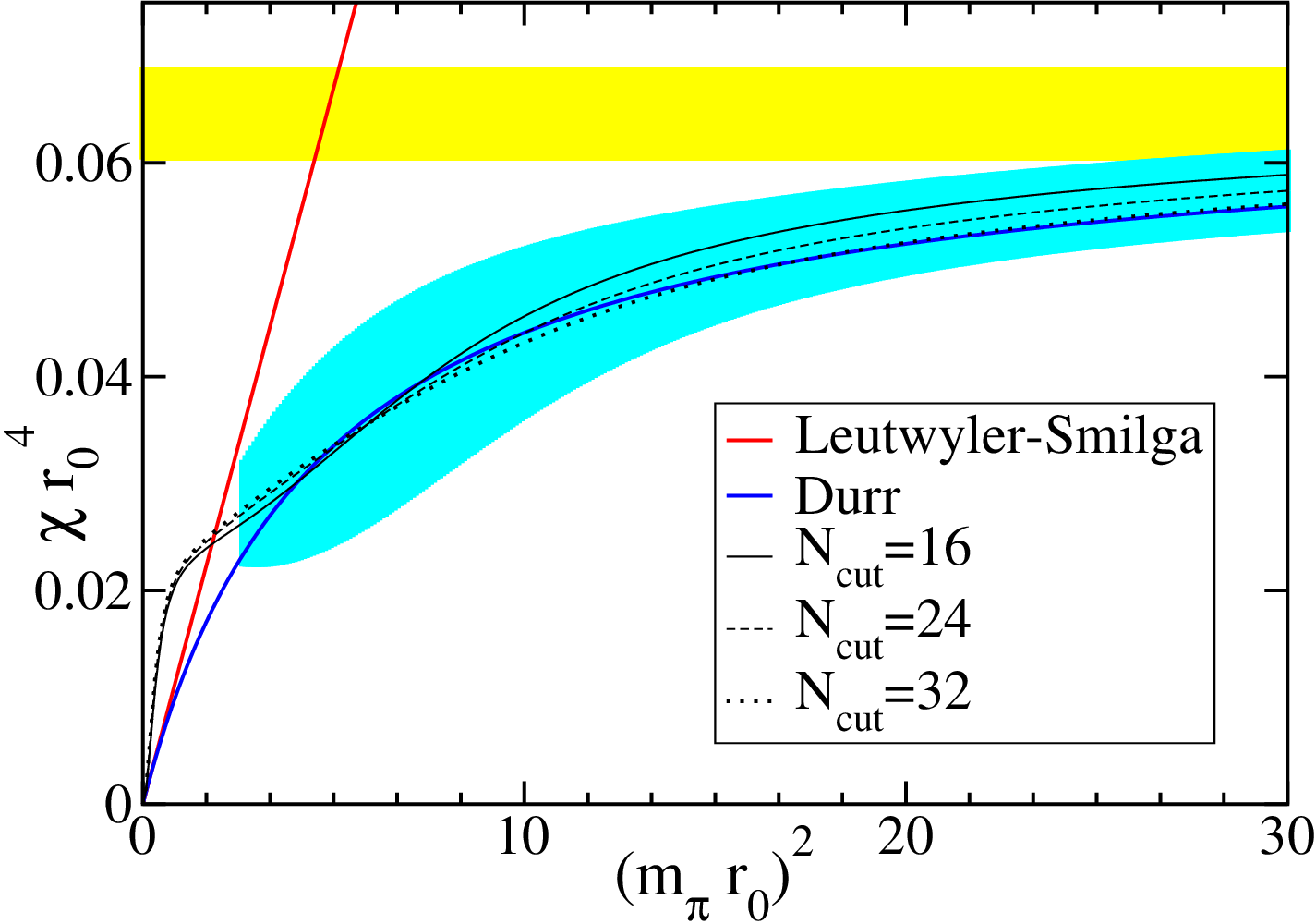}}
\vspace{-1.2cm}\caption{\label{fig:Nf2ov} The unquenched topological 
susceptibility vs. pion mass with overlap fermions.}
\end{minipage}
\vspace{-3mm}
         \label{fig:overlap}
\end{figure}
 
In Fig.\ \ref{fig:Nf2ov} I show the pion mass dependence of the 
topological susceptibility computed with truncations of $N=16,24,32$.
The steep line starting at the origin is the Leutwyler-Smilga prediction
(\ref{eq:LS}) and D\"urr's formula is also included using $f_\pi=93$MeV
and $\chi_{\mbox{\tiny quenched}}=(200$MeV$)^4$. The horizontal 
shaded region is the quenched estimate for the susceptibility from 
the same data set. One expects the
predictions based on broken chiral symmetry to be valid only if
$x \gg 1$. At the given volume the value of the parameter $x$ 
happens to roughly coincide with $(m_\pi r_0)^2$, so chiral 
symmetry is effectively broken if $(m_\pi r_0)^2 \gg 1$.
There are some indications that even in that regime, 
there are considerable finite volume
corrections to the unquenched susceptibility, much more so than in
the quenched case. A more detailed study of this question will 
appear in a separate publication.

\begin{figure}[htb!]
\begin{minipage}{75mm}
\resizebox{\textwidth}{!}{
\includegraphics{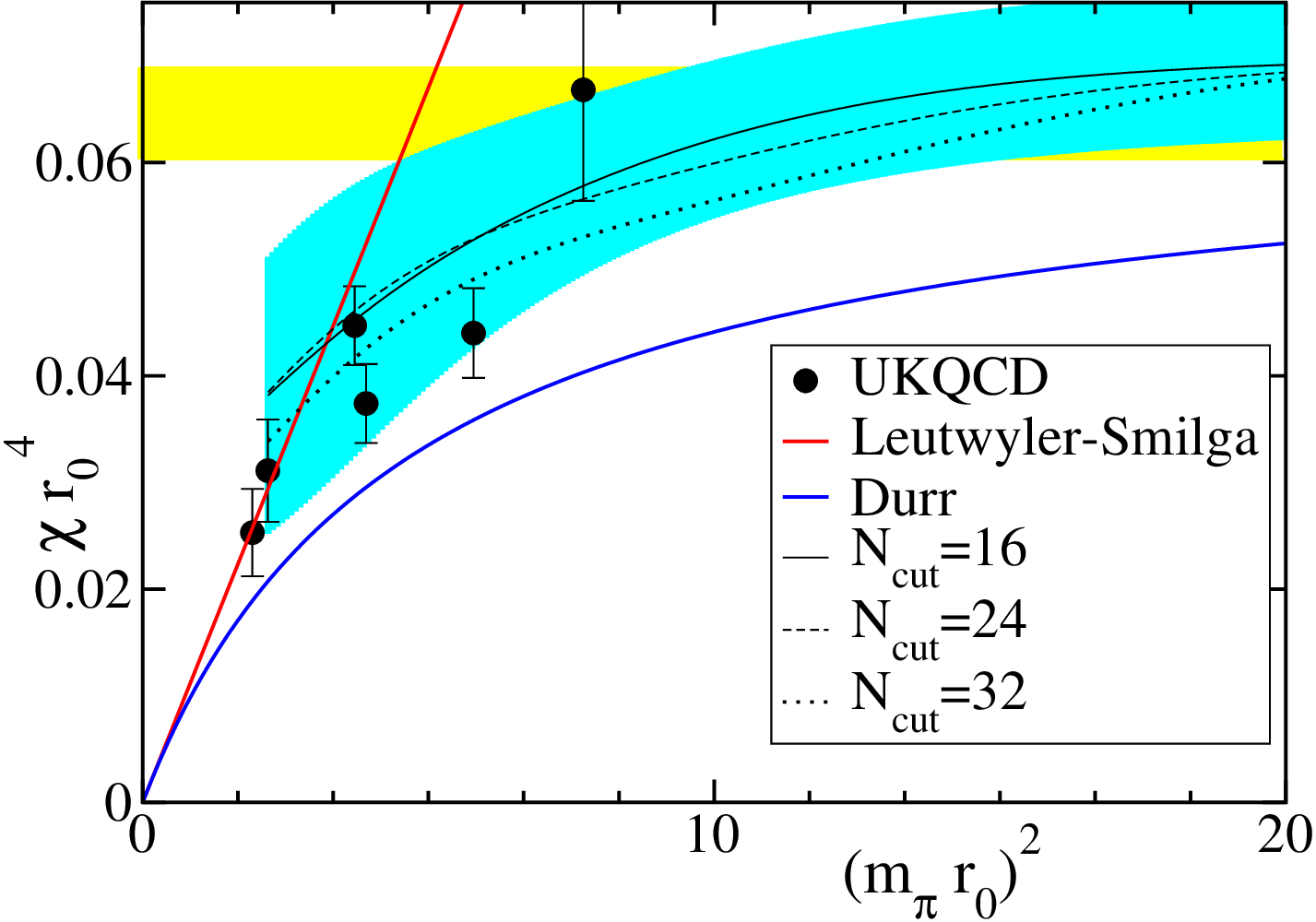}}
\vspace{-1.2cm}\caption{\label{fig:Nf2cl} The unquenched topological
susceptibility vs. pion mass with clover fermions.}
\end{minipage}
\vspace{-3mm}
\end{figure}

The same method can be used to compute the susceptibility with the
clover action. For this I take the non-perturbatively 
determined clover coefficient
$c_{sw}=1.91$ \cite{Edwards}. The results are shown in Fig.\ 
\ref{fig:Nf2cl} in a fashion similar to Fig.\ \ref{fig:Nf2ov}.
Here I also plotted the UKQCD results \cite{UKQCD} obtained 
with $N_f=2$ ${\cal O}(a)$ improved Wilson fermions. 

On the same set of configurations the clover action produces a 
significantly higher susceptibility than the overlap. This strongly
indicates that chirality of the fermion action can be important
for the topology. 
The error bars in the clover case are bigger than for the overlap, mainly due
to the weaker correlation between the topological sectors and the
fermionic determinant. This is even more pronounced in the case of
unimproved Wilson fermions (not shown in the Fig.), where this 
makes it impossible to obtain sensible results using reweighting and the
available statistics.


\begin{thebibliography}{9}

\bibitem{LS} H.\ Leutwyler and A.\ Smilga, Phys.\ Rev.\ D46 (1992) 5607.

\bibitem{Anna} A.\ Hasenfratz and C.\ Nieter, Phys.\ Lett.\ B439 (1998)
366.

\bibitem{Durr} S.\ D\"urr hep-lat/0103011 and hep-lat/0108015.

\bibitem{Hip} C.\ Gattringer and I.\ Hip, Nucl.\ Phys.\ B541 (1999) 305. 

\bibitem{Fodor} Z.\ Fodor and S.\ Katz, hep-lat/0104001 and hep-lat/0106002.

\bibitem{Smear} T.\ DeGrand, A.\ Hasenfratz and T.G.\ Kovacs, Nucl.\ Phys.\ 
B520 (1998) 301.

\bibitem{Edwards} R.G. Edwards, U.M. Heller, T.R. Klassen, Phys.\ Rev.\
Lett.\ 80 (1998) 3448.

\bibitem{UKQCD} A.\ Hart and M.\ Teper, hep-lat/0108022.

\end{thebibliography}
\end{document}